\documentclass[fleqn,twoside]{article}
\usepackage{espcrc2}
\usepackage{graphics}

\def\Vec#1{\mbox{\boldmath $#1$}}

\title{
Search for the Electric Dipole Moment of the $\tau$ Lepton
}

\author{
K.Inami
\address[Nagoya]{Department of Physics, Nagoya University, \\
Furo-cho, Chikusa-ku, Nagoya, 464-8602, Japan}
for the Belle collaboration
}

\begin{document}

\begin{abstract}
We have searched for a T/CP violation signature arising from
an electric dipole form factor ($d_\tau$) of the $\tau$ lepton
in the $e^+e^- \to \tau^+\tau^-$ reaction.
Using an optimal observable method
for 29.5 fb$^{-1}$ of data collected with the Belle detector 
at the KEKB collider at $\sqrt{s}=10.58$~GeV,
we obtained the preliminary result
$Re(d_\tau) = ( 1.15 \pm 1.70 ) \times 10^{-17} e{\rm cm}$ and
$Im(d_\tau) = ( -0.83 \pm 0.86 ) \times 10^{-17} e{\rm cm}$
for the real and imaginary parts of $d_\tau$, respectively,
and set the 95\% confidence level limits
$-2.2 < Re(d_\tau) < 4.5 ~(10^{-17}e{\rm cm})$ and
$-2.5 < Im(d_\tau) < 0.8 ~(10^{-17}e{\rm cm})$.
\end{abstract}

\maketitle

\section{Introduction}

While large CP violating asymmetry in the quark sector 
has recently been confirmed in B-meson decay at B 
factories~\cite{ref:BCPV1,ref:BCPV2}, 
the Standard Model (SM) does not predict
any appreciable CP violation (CPV) in the lepton sector.
However, physics beyond the SM, such as multi-Higgs, SUSY, and
lepto-quark models~\cite{ref:th1,ref:th2},
could result in CPV
in leptonic processes; since new bosons and Higgs would strongly couple
with heavy particles through quantum loop effects, CP violating
phenomena could be significantly enhanced, particularly in $\tau$ decay, due 
to its very large mass compared to other leptons.

The contribution from CP violating interactions in the $\tau$-pair
production process can be parametrized
in a model-independent way using the electric dipole moment
(or form factor) for the $\tau$ lepton.
Up to now, searches for the electric dipole form factor $d_{\tau}$
have been performed at LEP~\cite{ref:L3,ref:OPAL} in the reaction $e^+e^- \to 
\tau^+\tau^-\gamma$ and by ARGUS~\cite{ref:ARGUS} in the $e^+e^- \to 
\tau^+\tau^-$ reaction
and the following
upper limits at 95\% confidence level
were obtained: $-3.1<d_\tau<3.1$ ($10^{-16}e$cm) from L3~\cite{ref:L3},
$-3.8<d_\tau<3.6$ ($10^{-16}e$cm) from OPAL~\cite{ref:OPAL}
and $|Re(d_\tau)|<4.6\times10^{-16}e$cm, $|Im(d_\tau)|<1.8\times10^{-16}e$cm
from ARGUS~\cite{ref:ARGUS}. \\

We have searched for a form factor $d_\tau$
at the $\gamma\tau\tau$ vertex 
in the process $e^+e^- \to \gamma^* \to \tau^+ \tau^-$
using triple momentum and spin correlation observables. 
The effective Lagrangian can be expressed as
\begin{equation}
{\cal L}_{CP} = - i d_\tau(s) \bar{\tau} \sigma^{\mu \nu} \gamma_5 \tau
  \partial_\mu A_\nu, \label{eq:Lagrangian}
\end{equation}
where the electric dipole form factor $d_\tau$ depends in general on $s$,
the squared energy of the $\tau$-pair system. 
In common with other analyses we ignore this possible $s$-dependence,
assuming $d_\tau(s) \equiv d_\tau$, constant.
The squared spin density matrix (${\cal M}^2_{\rm prod}$) for the reaction 
$e^+(\Vec{p})e^-(-\Vec{p}) \to \tau^+(\Vec{k},\Vec{S}_+) \tau^-(-\Vec{k},
\Vec{S}_-)$ is given by~\cite{ref:EDM}
\small
\begin{eqnarray}
& & \!\!\!\!\!\!\!\!\!\!\!\!\!\!
{\cal M}_{\rm prod}^2 = {\cal M}_{\rm SM}^2 
 + Re(d_\tau) {\cal M}_{Re}^2
 + Im(d_\tau) {\cal M}_{Im}^2, \\
& & \!\!\!\!\!\!\!\!\!\!\!\!\!\!
{\cal M}_{\rm SM}^2 = 
\nonumber \\ 
& & \!\!\!\!\!\!\!\!\!\!\!\!\!\! ~~
\frac{e^4}{k_0^2}
[   k_0^2 + m_\tau^2 + |\Vec{k}^2|(\hat{\Vec{k}}\hat{\Vec{p}})^2 
 - \Vec{S}_+ \Vec{S}_- |\Vec{k}|^2 (1-(\hat{\Vec{k}}\hat{\Vec{p}})^2) ]
\nonumber \\ 
& & \!\!\!\!\!\!\!\!\!\!\!\!\!\! ~~
 + 2(\hat{\Vec{k}}\Vec{S}_+)(\hat{\Vec{k}}\Vec{S}_-)
    (|\Vec{k}|^2+(k_0 - m_\tau)^2 (\hat{\Vec{k}}\hat{\Vec{p}})^2)
\nonumber \\ 
& & \!\!\!\!\!\!\!\!\!\!\!\!\!\! ~~
 - 2 k_0(k_0-m_\tau)(\hat{\Vec{k}}\hat{\Vec{p}})
    ((\hat{\Vec{k}}\Vec{S}_+)(\hat{\Vec{p}}\Vec{S}_-) 
    +(\hat{\Vec{k}}\Vec{S}_-)(\hat{\Vec{p}}\Vec{S}_+) )
\nonumber \\ 
& & \!\!\!\!\!\!\!\!\!\!\!\!\!\! ~~
 + 2 k_0^2 (\hat{\Vec{p}}\Vec{S}_+) (\hat{\Vec{p}}\Vec{S}_-),
\\
& & \!\!\!\!\!\!\!\!\!\!\!\!\!\!
{\cal M}_{Re}^2 = 4 \frac{e^3}{k_0} |\Vec{k}| [
 - (m_\tau + (k_0-m_\tau)(\hat{\Vec{k}}\hat{\Vec{p}})^2)
   (\Vec{S}_+ \times \Vec{S}_-)\hat{\Vec{k}}
\nonumber \\ 
& & \!\!\!\!\!\!\!\!\!\!\!\!\!\! ~~
 + k_0 (\hat{\Vec{k}}\hat{\Vec{p}})(\Vec{S}_+ \times \Vec{S}_-)\hat{\Vec{p}}
],
\\
& & \!\!\!\!\!\!\!\!\!\!\!\!\!\!
{\cal M}_{Im}^2 = 4 \frac{e^3}{k_0} |\Vec{k}| [
 - (m_\tau + (k_0-m_\tau)(\hat{\Vec{k}}\hat{\Vec{p}})^2)
   (\Vec{S}_+ - \Vec{S}_-)\hat{\Vec{k}}
\nonumber \\ 
& & \!\!\!\!\!\!\!\!\!\!\!\!\!\! ~~
 + k_0 (\hat{\Vec{k}}\hat{\Vec{p}})(\Vec{S}_+ - \Vec{S}_-)\hat{\Vec{p}}
],
\end{eqnarray}
\normalsize
where $k_0$ is the energy of the $\tau$, $m_\tau$ is the $\tau$ mass, 
$\Vec{p}$ is the momentum vector of $e^+$, $\Vec{k}$ is 
the momentum vector of $\tau^+$ in the center-of-mass frame,
$\Vec{S}_\pm$ are the spin vectors for $\tau^\pm$, and the hat denotes 
a unit momentum.
We disregard the higher order terms proportional to $|d_\tau^2|$
assuming $d_\tau(s)$ to be small.
${\cal M}_{\rm SM}^2$ corresponds to the SM term.
${\cal M}_{Re}^2$ and ${\cal M}_{Im}^2$ are the interference terms 
(related to the real and imaginary parts of $d_\tau$, respectively)
between the SM and CPV amplitudes.
${\cal M}_{Re}^2$ is CP odd and T odd, while ${\cal M}_{Im}^2$ is CP
odd, but T even.
In the above equations, $e^+$ and $e^-$ are assumed to be unpolarized and
massless particles.

We adapt the so-called optimal observable method~\cite{ref:Optimal}, which 
maximizes the sensitivity to $d_\tau$  by optimizing the relevant 
quantities, ${\cal O}_{Re}$ and ${\cal O}_{Im}$, for measured kinematic 
variables, as described below. Here the optimal observables are defined as
\begin{equation}
{\cal O}_{Re} = \frac{{\cal M}_{Re}^2}{{\cal M}_{\rm SM}^2},~~~
{\cal O}_{Im} = \frac{{\cal M}_{Im}^2}{{\cal M}_{\rm SM}^2}.
\end{equation}
The mean value of the observables is expressed as 
\begin{eqnarray}
\langle{\cal {O}}_{Re}\rangle &\!\!\!\!\!\propto&\!\!\!\!\!
\int {\cal O}_{Re} d\sigma \propto
\int {\cal O}_{Re} {\cal M}^2_{\rm prod} d\phi
\nonumber \\
&\!\!\!\!\!=&\!\!\!\!\! \int {\cal{M}}^2_{Re} d\phi + 
Re(d_{\tau}) \int \frac{({\cal {M}}^2_{Re})^2}
{{\cal {M}}^2_{\rm{SM}}} d\phi, ~~~~~
\label{eq:obs}
\end{eqnarray}
where the integration is over the phase space ($\phi$) sustained by 
the relevant kinematic variables.
The cross-term containing the integral of the product of
${\cal M}^2_{Re}$ and ${\cal M}^2_{Im}$ drops out 
because of their different symmetry properties.
The expression for the imaginary part is similar.
The means of the observables
$\langle {\cal O}_{Re} \rangle$ and $\langle {\cal O}_{Im} \rangle$ 
are thus expressed as linear functions of 
$d_{\tau}$
\begin{eqnarray}
 \langle{\cal O}_{Re}\rangle &\!\!\!\!\!=&\!\!\!\!\!
   a_{Re} \cdot Re(d_\tau) +b_{Re},
\nonumber \\
 \langle{\cal O}_{Im}\rangle &\!\!\!\!\!=&\!\!\!\!\!
   a_{Im} \cdot Im(d_\tau) +b_{Im}.
 \label{eq:relation1}
\end{eqnarray}

Eight different final states in the decay of $\tau$-pairs,
$(e\nu\bar{\nu})(\mu\nu\bar{\nu})$, $(e\nu\bar{\nu})(\pi\nu)$, 
$(\mu\nu\bar{\nu})(\pi\nu)$, $(e\nu\bar{\nu})(\rho\nu)$, 
$(\mu\nu\bar{\nu})(\rho\nu)$, $(\pi\nu)(\rho\nu)$, 
$(\rho\nu)(\rho\bar{\nu})$, and $(\pi\nu)(\pi\bar{\nu})$, are analyzed,
where all particles except $\nu$ and $\bar{\nu}$ are positively or negatively 
charged.
Because of the undetectable particles, we can not fully reconstruct
the quantities $\Vec{k}$ and $\Vec{S}_\pm$. 
Therefore, for each event we calculate possible kinematic 
configurations and obtain 
the mean value of ${\cal M}^2_{\rm SM}$, 
${\cal M}^2_{Re}$ and ${\cal M}^2_{Im}$
by averaging over the calculated configurations.
In the case when both $\tau$ leptons decay hadronically 
($\tau \to \pi\nu$ or $\rho\nu$),
the $\tau$ flight direction is calculated with a two-fold ambiguity 
and we take the average of ${\cal M}^2_{\rm SM}$, 
${\cal M}^2_{Re}$ and ${\cal M}^2_{Im}$ over the two solutions.
In the case when either one or both $\tau$ leptons decay leptonically
($\tau \to e\nu\bar{\nu}$ or $\mu\nu\bar{\nu}$), 
a Monte Carlo (MC) treatment is adopted to take into account
the additional ambiguity in the effective mass of the $\nu\bar{\nu}$ system 
($m_{\nu\bar{\nu}}$). 
For each event we generate 100 MC configurations using a hit-and-miss approach
by varying $m_{\nu\bar{\nu}}$, and compute the averaged 
${\cal M}^2_{\rm SM}$, ${\cal M}^2_{Re}$ and ${\cal M}^2_{Im}$
over successful tries
in which the $\tau$ direction can be constructed kinematically. 
In the calculation,
we ignore the effect of undetected photons coming from initial
state radiation, radiative $\tau$ decays and bremsstrahlung.
This effect has been examined and found to be negligible
compared to the total errors.

\section{Data and event selection}

In this analysis, we used 26.8 million $\tau$-pairs ($29.5~{\rm fb}^{-1}$)
accumulated with 
the Belle detector~\cite{ref:Belle} at the KEKB accelerator~\cite{ref:KEKB}.
KEKB is an asymmetric energy $e^+e^-$ collider
with a beam crossing angle of 22 mrad. Its center-of-mass energy is
10.58 GeV, corresponding to the $\Upsilon(4S)$ resonance, 
with beam energies of 8 and 3.5 GeV for electrons and positrons,
respectively.
Belle is a general-purpose detector with an asymmetric structure 
along the beam direction. Among the detector elements, 
the central drift chamber (CDC) and the silicon vertex detector (SVD)
are essential to obtain the momentum vectors of charged particles.
The combined information from the silica Aerogel Cherenkov counters (ACC),
the time-of-flight counters (TOF), the CsI electromagnetic calorimeter (ECL), 
and the $\mu/{\rm K_L}$ detector (KLM) is used for particle identification.

The MC event generators
KORALB / TAUOLA~\cite{ref:KORALB} are used for 
$\tau$-pair production and decays.
The detector simulation is performed by a GEANT-based program, GSIM.
Actual data and MC generated events are reconstructed 
by the same program written by the Belle collaboration. \\

\begin{table*}[htb]
\renewcommand {\baselinestretch}{0.75}
 \caption{Yield, purity and background rate obtained by the event selection
described in the text, where the purity was evaluated by MC
simulation and its error comes from the MC statistics. }
 \label{table:selection.result}
 \begin{tabular}{crcl}
  \hline
       & Yield    & Purity (\%)     & Background mode (\%)\\
  \hline
  $e\mu$     & 250,948    & $96.6\pm0.1$ & $2\gamma \to \mu\mu$(1.9), $\tau\tau \to e\pi$(1.1). \\
  $e\pi$     & 132,574    & $82.5\pm0.1$ & $\tau\tau \to e\rho$(6.0), $eK$(5.4), $e\mu$(3.1), $eK^*$(1.3). \\
  $\mu\pi$   & 123,520    & $80.6\pm0.1$ & $\tau\tau \to \mu\rho$(5.7), $\mu K$(5.3), $\mu\mu$(2.9), $2\gamma \to \mu\mu$(2.0). \\
  $e\rho$    & 240,501    & $92.4\pm0.1$ & $\tau\tau \to e\pi\pi^0\pi^0$(4.4), $eK^*$(1.7). \\
  $\mu\rho$  & 217,156    & $91.6\pm0.1$ & $\tau\tau \to \mu\pi\pi^0\pi^0$(4.2), $\mu K^*$(1.6), $\pi\rho$(1.0). \\
  $\pi\rho$  & 110,414    & $77.7\pm0.1$ & $\tau\tau \to \rho\rho$(5.1), $K \rho$(4.9), $\pi\pi\pi^0\pi^0$(3.8), $\mu\rho$(2.7). \\
  $\rho\rho$ & 93,016     & $86.2\pm0.1$ & $\tau\tau \to \rho\pi\pi^0\pi^0$(8.0), $\rho K^*$(3.1). \\
  $\pi\pi$   & 28,348     & $70.0\pm0.2$ & $\tau\tau \to \pi\rho$(9.2), $\pi K$(9.2), $\pi\mu$(4.7), $\pi K^*$(2.0). \\
  \hline
 \end{tabular}
\end{table*}

We chose the eight final state $\tau$-pair modes mentioned above.
All of the final state particles were reconstructed with the following 
conditions.
Each charged track is required to have a transverse momentum $p_t>0.1$ 
GeV/$c$. 
Photon candidates should deposit an energy of $E>0.1$ GeV 
in the ECL.
A signal event is required to have two charged tracks with zero net-charge 
and no photon apart from $\rho^\pm \to  \pi^\pm \pi^0$, 
$\pi^0 \to \gamma \gamma$.

A track was identified as an electron using a likelihood ratio
combining $dE/dx$ in the CDC,
the ratio of energy deposited in the ECL and momentum
measured in the CDC,
the shower shape of the ECL
and the hit pattern from the ACC.
The identification efficiency was estimated to be 92\% 
with a $\pi^\pm$ fake rate of 0.3\%~\cite{ref:eID}.
A muon was evaluated from its range and hit pattern at the KLM detector; 
its efficiency and fake rate were obtained by MC as 91\% and 2\%,
respectively.
A pion track was found by requiring
that a track be identified as a hadron by the KLM information,
and not identified as an electron.
The efficiency was estimated to be 81\%.
The purity for the obtained samples is about 89\%.
A $\rho^\pm$ was reconstructed from a charged track and a $\pi^0$ where 
the track should be neither an electron nor a muon,
and for $\pi^0 \to \gamma\gamma$, the reconstructed $\pi^0$ should have an 
invariant mass between 110 and 150 MeV/$c^2$ and 
a momentum in the laboratory frame larger than $0.2$ GeV/$c$.
In order to suppress the background
and to enhance the particle identification ability,
the detection of leptonic particles was restricted to within 
the barrel region, $-0.60<\cos \theta^{\rm lab} < 0.83$,
and that of single pions to within the KLM barrel region,
$-0.50<\cos \theta^{\rm lab}<0.62$,
where $\theta^{\rm lab}$ is the polar angle 
relative to the $e^-$ beam direction.
For the same reason, the particle momentum was required to be greater 
than 0.5 GeV/$c$ for an electron, 1.2 GeV/$c$ for both a muon and pion,
and 1.0 GeV/$c$ for $\rho^\pm$ in the laboratory frame.

The dominant backgrounds are due to two-photon as well as Bhabha and 
$\mu\mu$ processes. 
In order to remove two-photon events,
we required the missing momentum not to be directed towards the beam-pipe
region 
(imposing a selection $-0.950 < \cos \theta^{\rm lab}<0.985$),
and to reject the latter processes we required
that the sum of the charged track momenta
be less than 9 GeV/$c$ in the center-of-mass frame.
Additional selections were imposed particularly on the $e\pi$ mode 
where a large number of Bhabha events could contribute
through misidentification. For the $e\pi$ mode,
we remove events which satisfy the following criteria:
the opening angle of the two tracks in the plane perpendicular to the beam axis
is greater than $175^\circ$, and their momentum sum is greater
than 6 GeV/$c$ in the $\tau$-pair rest frame.
Finally, we removed events in which
the $\tau$ flight direction could not be kinematically reconstructed,
which mostly arise from $\tau$-pairs 
having hard initial-state radiation 
and misidentified $\tau$-pair backgrounds.

The yield of events passing this selection is given in 
Table~\ref{table:selection.result} for each of the eight selected modes.
The mean energy of the $\tau$-pair system in the obtained sample
is $\sqrt{s}=10.38$~GeV; this sets the scale at which $d_\tau(s)$ is measured.
Because of events with soft radiated photons,
the energy scale is slightly lower than the beam energy.
The dominant background sources are also listed in the table.
Hadronic $\tau$ decays with two or more $\pi^0$'s make a large contribution
of a few percent.
For the modes including $\pi^\pm$, the misidentification of kaons and muons
as pions yields other backgrounds;
for example, 5.3\% of $\mu K$ final states is included in the $\mu \pi$ mode.
The other backgrounds are estimated by MC to be a few percent
from two-photon processes,
and less than 1\% from Bhabha, $\mu\mu$, and multihadronic processes.

\begin{figure}[tb]
\centerline{\resizebox{3.7cm}{3.7cm}{\includegraphics{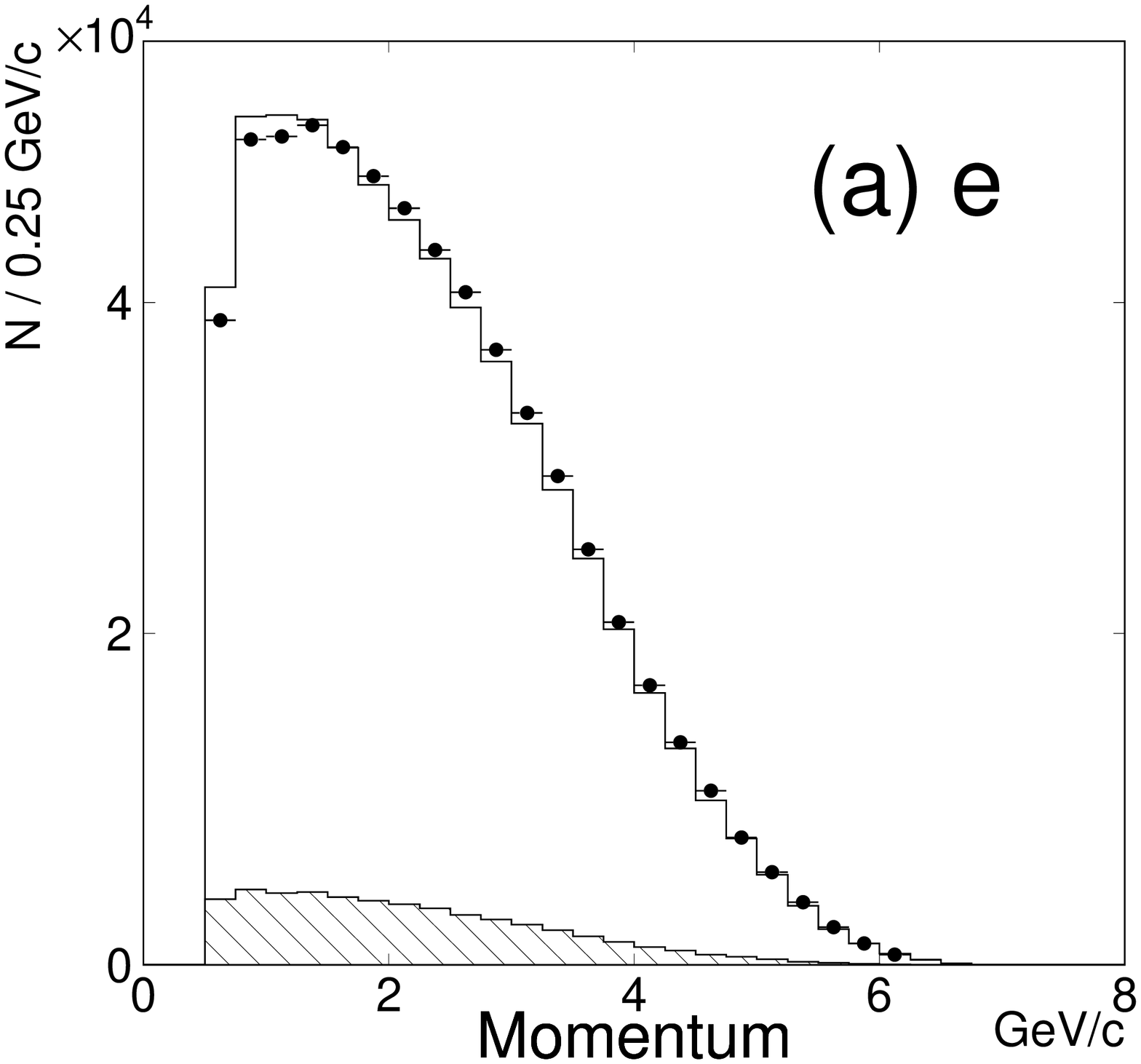}}
            \resizebox{3.7cm}{3.7cm}{\includegraphics{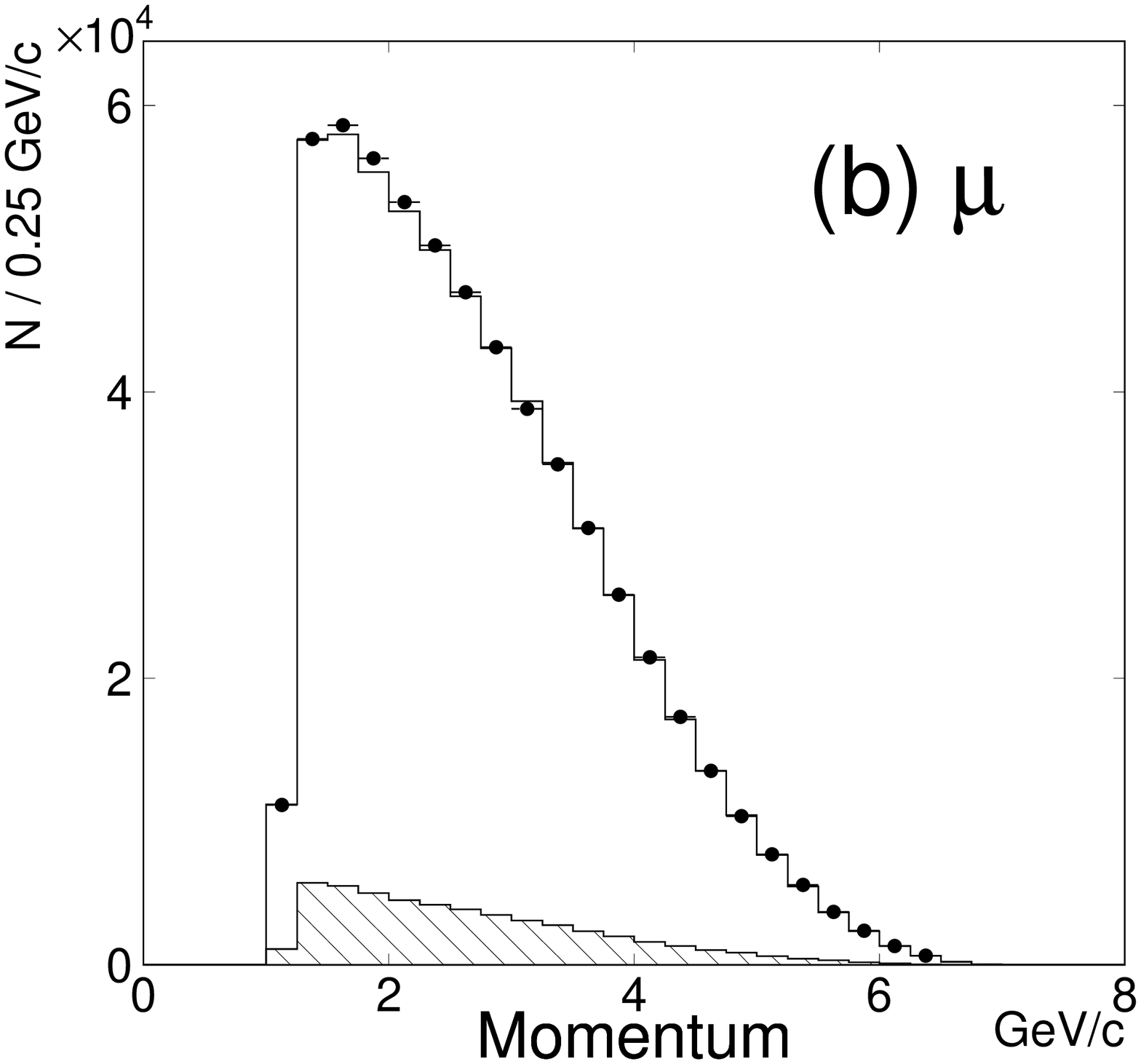}}}
\centerline{\resizebox{3.7cm}{3.7cm}{\includegraphics{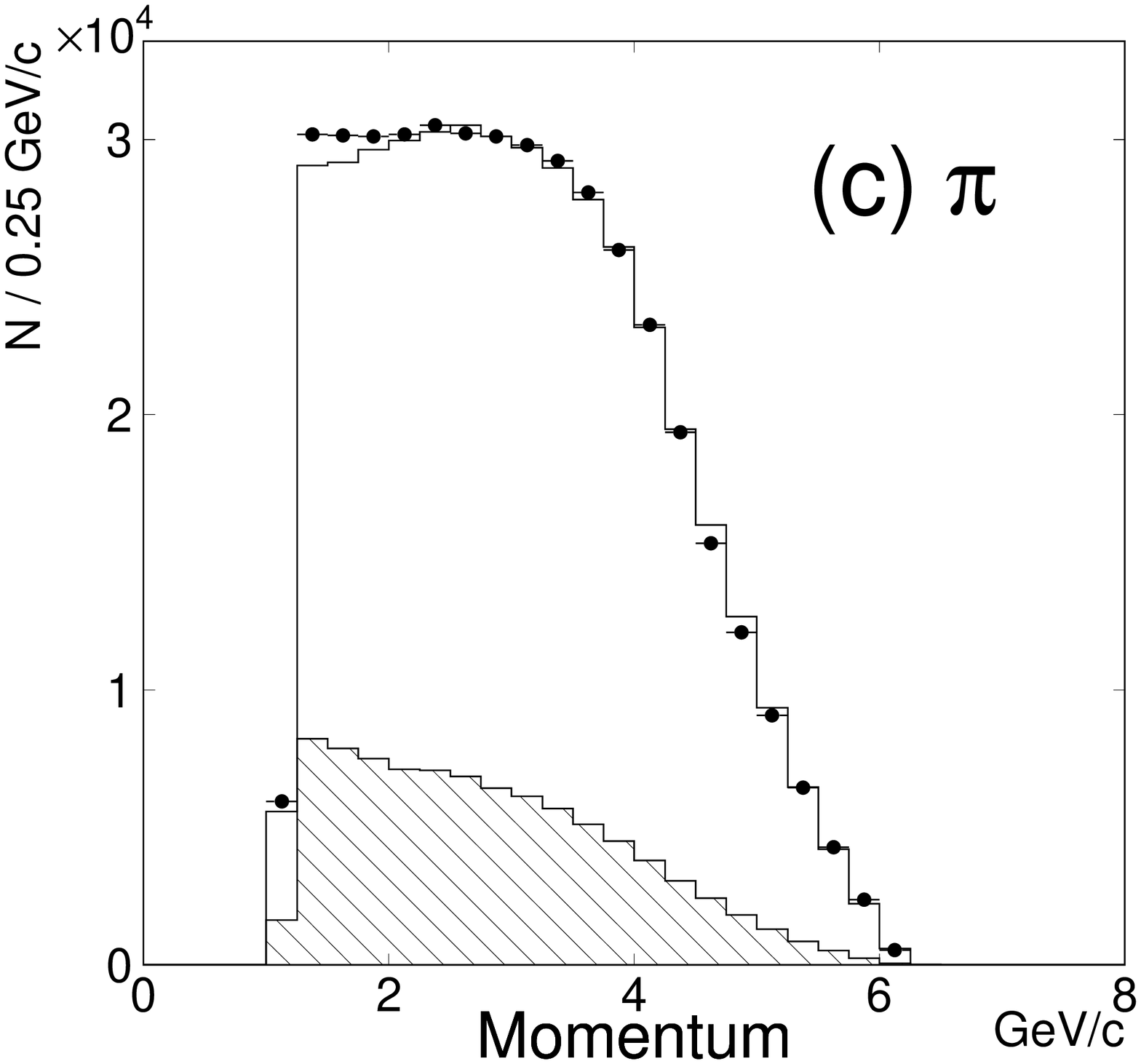}}
            \resizebox{3.7cm}{3.7cm}{\includegraphics{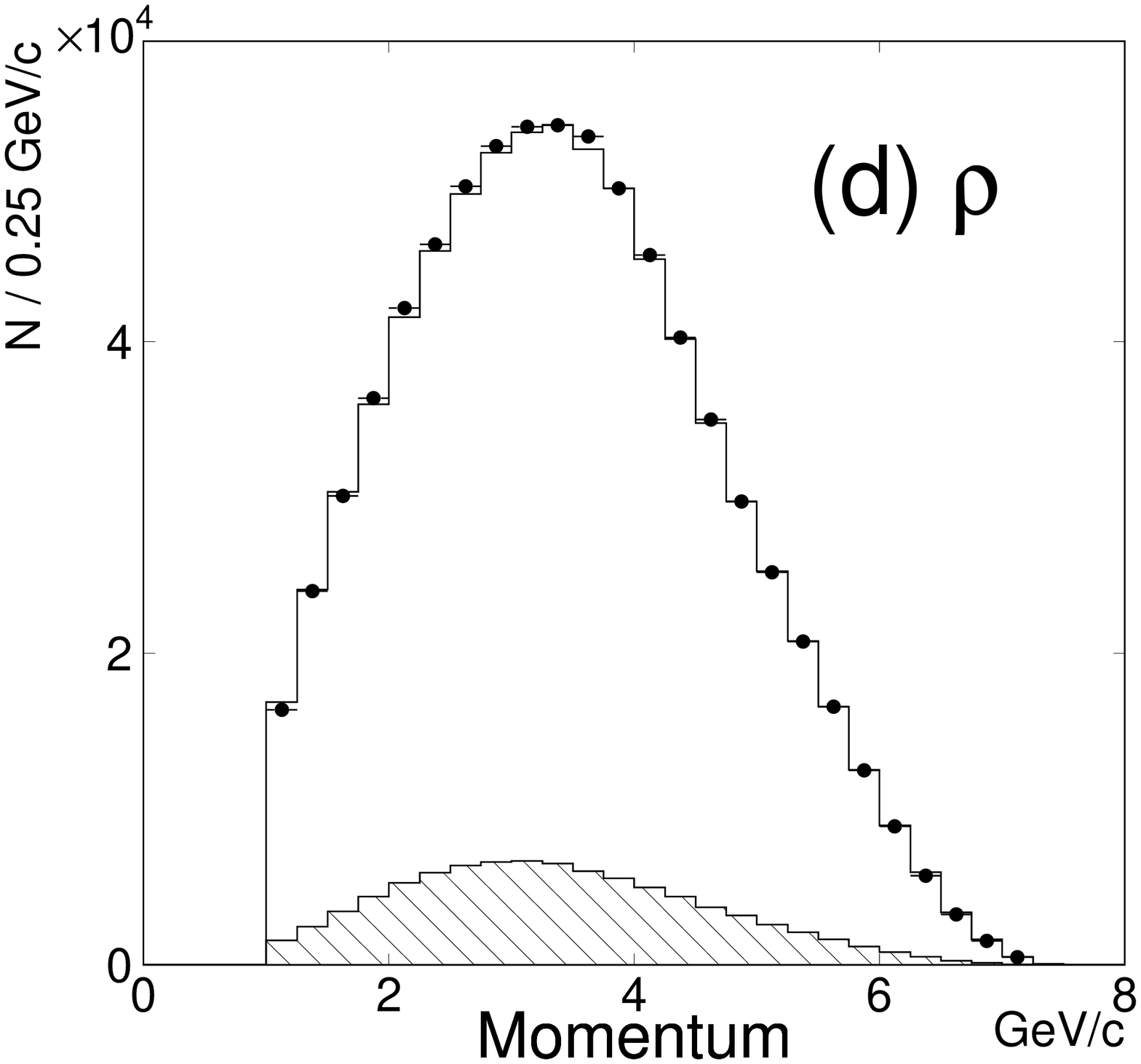}}}
\renewcommand {\baselinestretch}{0.75}
\caption{Momentum distributions of 
(a) $e^\pm$, (b) $\mu^\pm$, (c) $\pi^\pm$, and (d) $\rho^\pm$ in 
the laboratory frame.
The points with error bars are the data and the histogram is the MC 
expectation. 
The latter is scaled to the total number of entries.
The hatched histogram is the background distribution evaluated by MC.}
\label{fig:mom}
\end{figure}

\begin{figure}[tb]
\centerline{\resizebox{3.7cm}{3.7cm}{\includegraphics{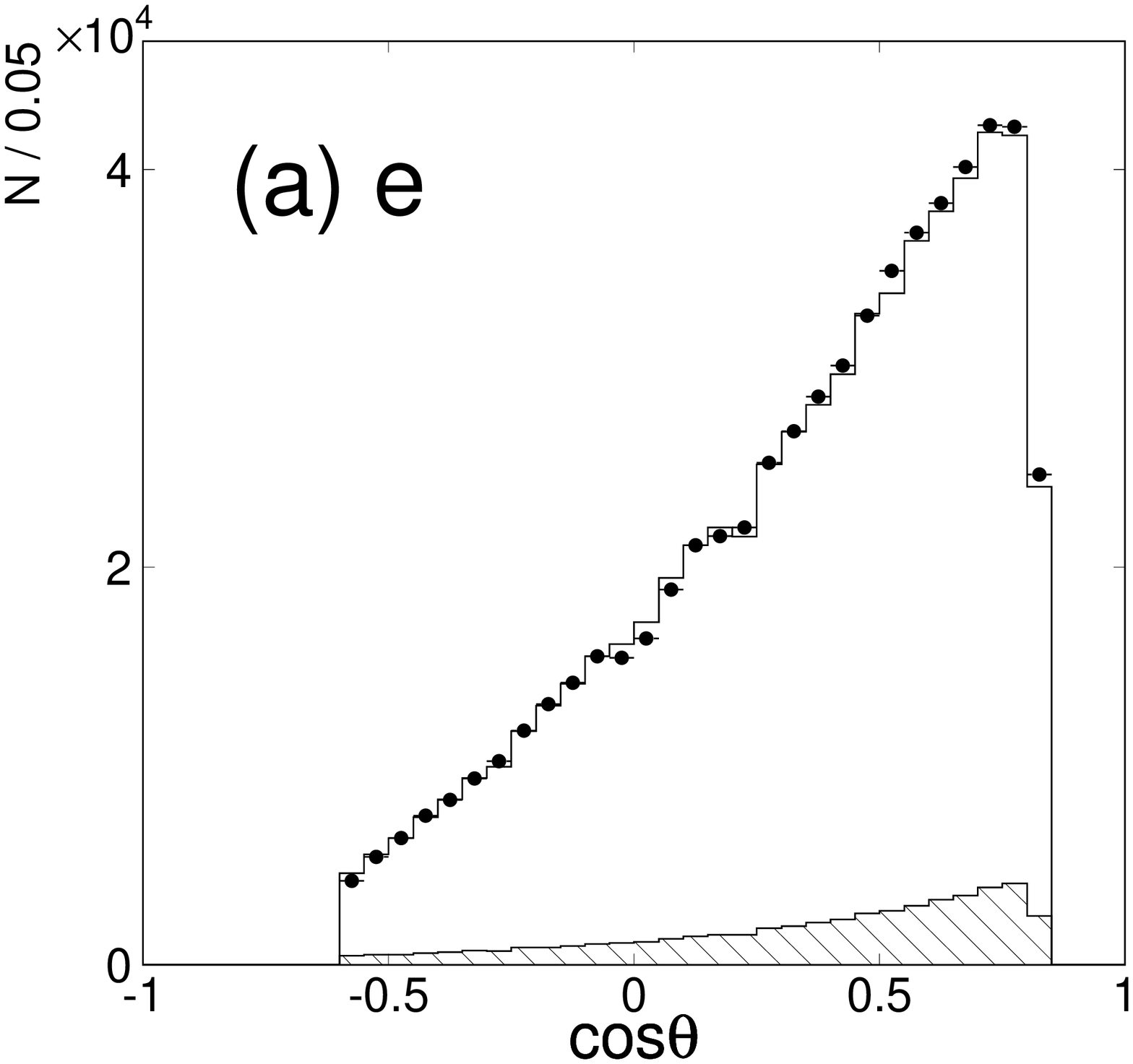}}
            \resizebox{3.7cm}{3.7cm}{\includegraphics{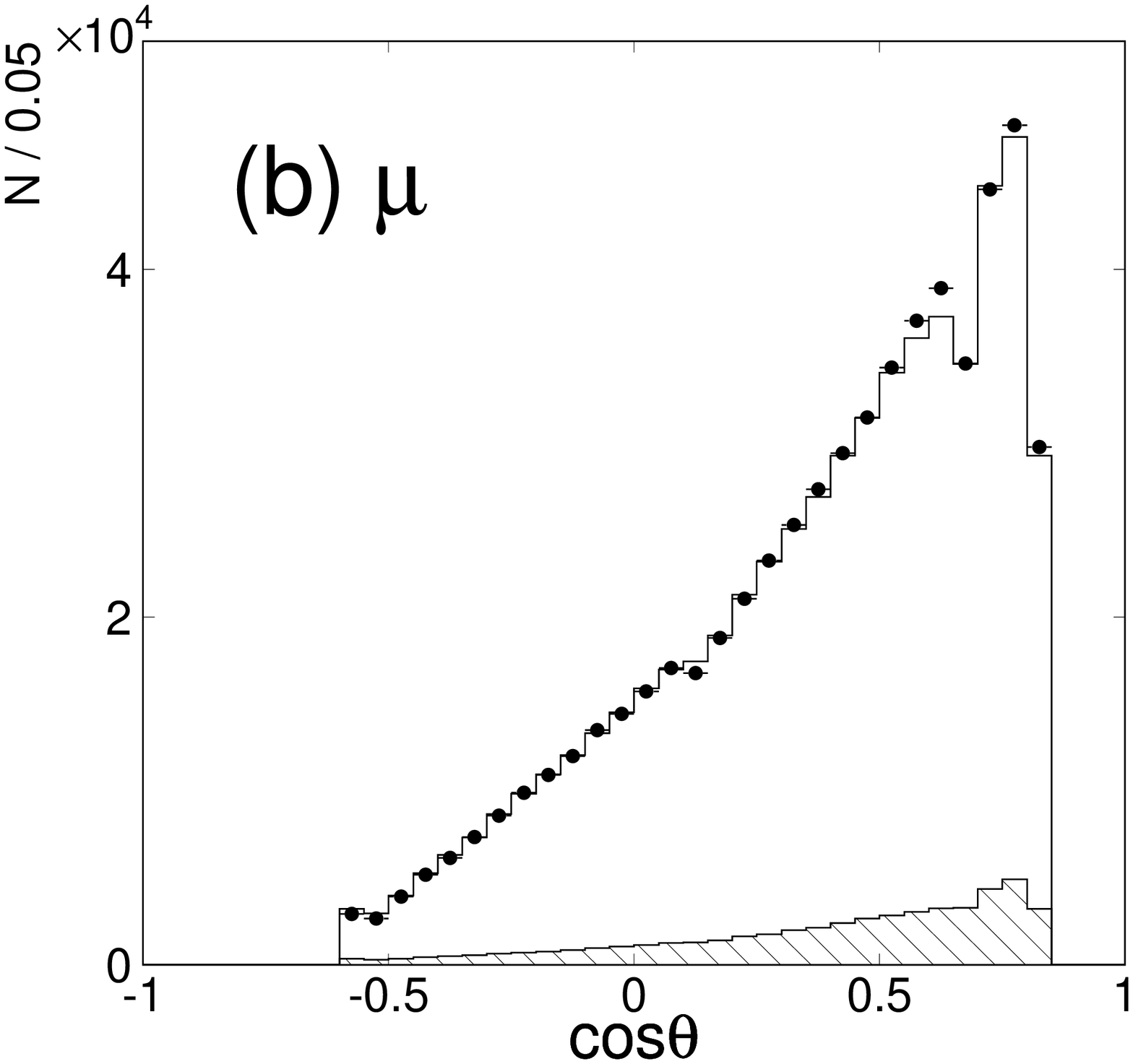}}}
\centerline{\resizebox{3.7cm}{3.7cm}{\includegraphics{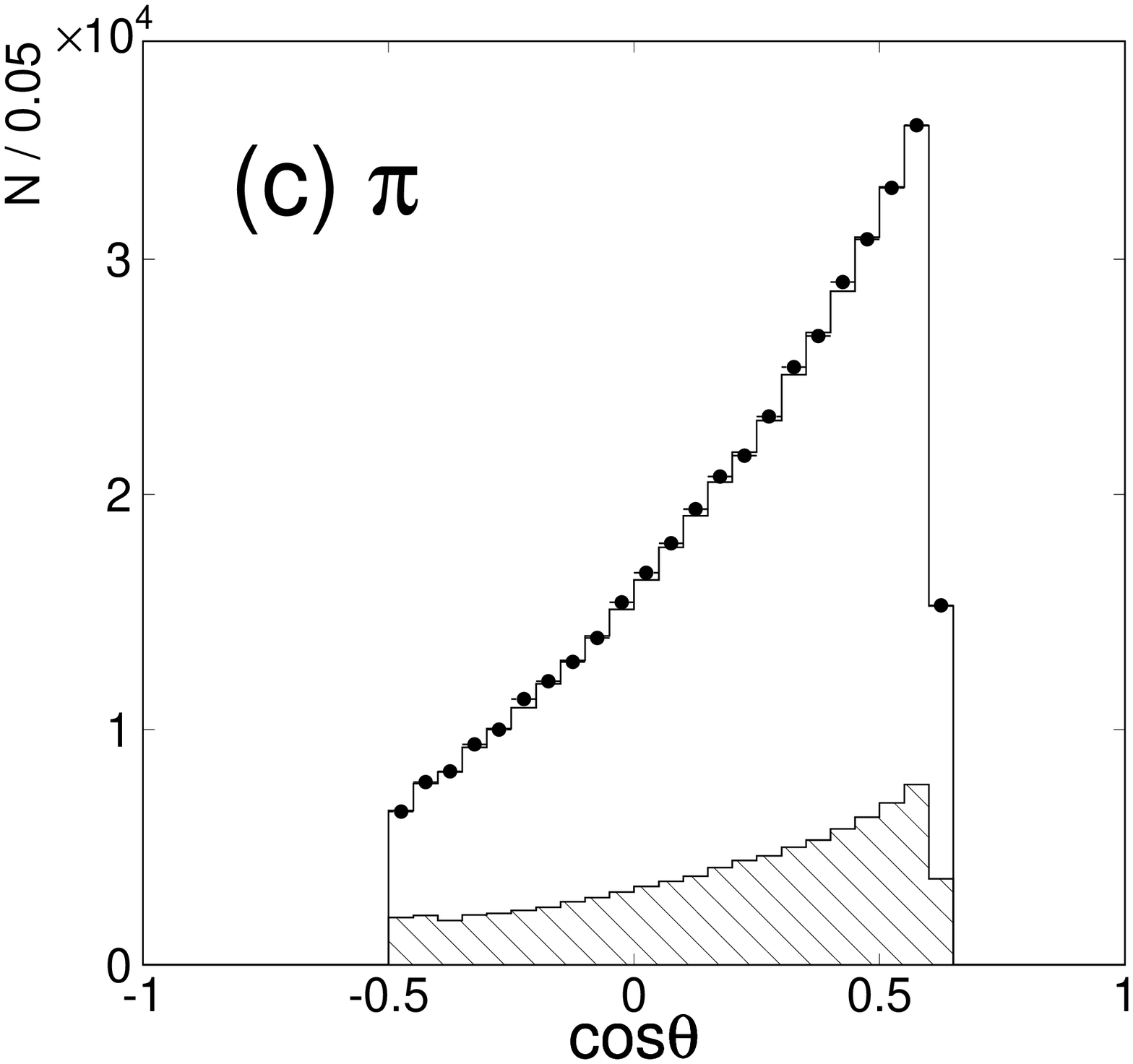}}
            \resizebox{3.7cm}{3.7cm}{\includegraphics{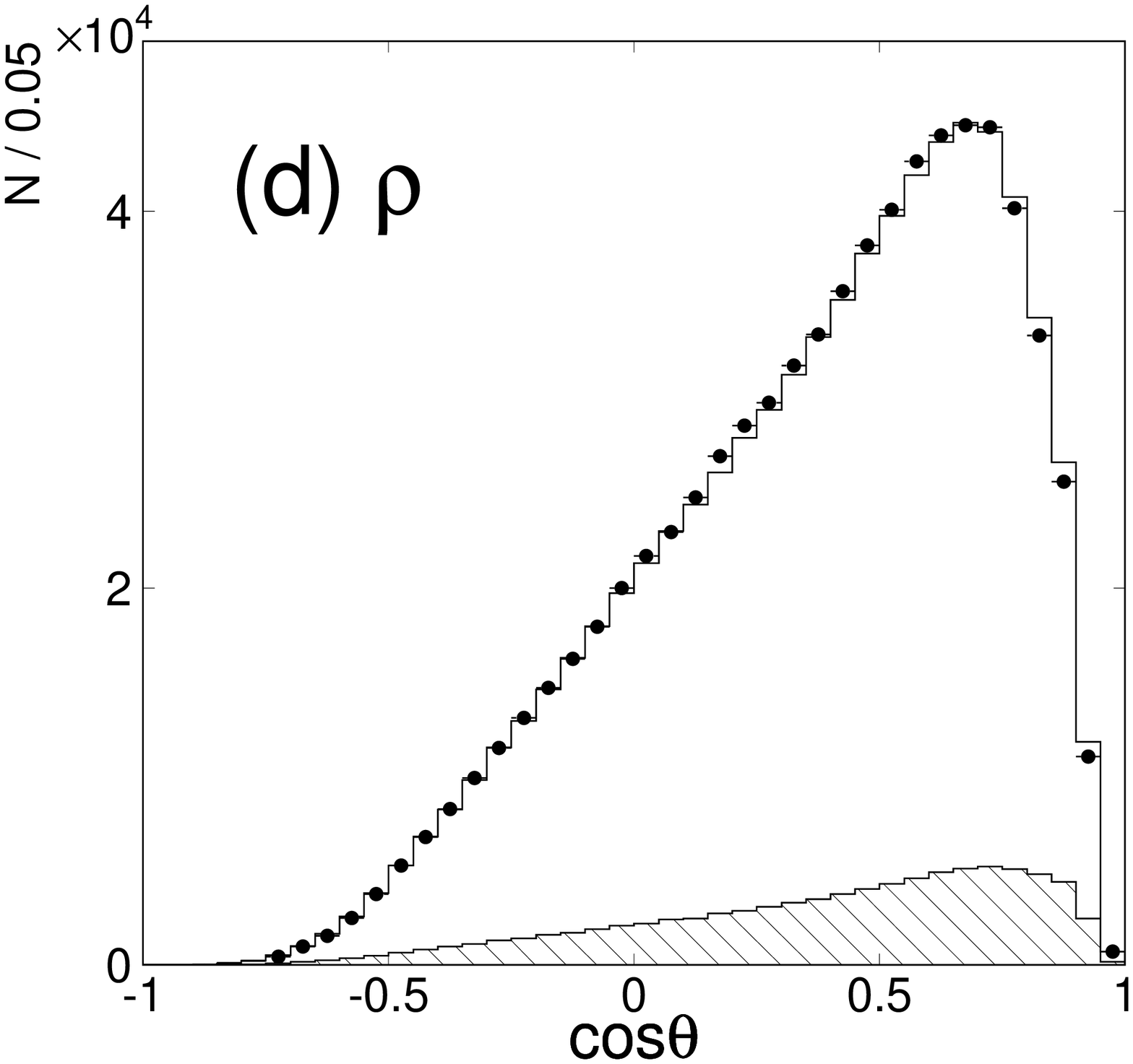}}}
\renewcommand {\baselinestretch}{0.75}
\caption{The $\cos \theta$ distributions of
(a) $e^\pm$, (b) $\mu^\pm$, (c) $\pi^\pm$, and (d) $\rho^\pm$ in 
the laboratory frame.
The meanings of the points and histograms are the same as 
in Fig.~\ref{fig:mom}.}
\label{fig:cos}
\end{figure}

Figs~\ref{fig:mom} and \ref{fig:cos} show the resulting 
momentum and $\cos \theta^{lab}$ distributions, respectively, 
for charged particles in the laboratory frame.
Very good agreement with MC is found,
except for low-momentum electrons (Fig.~\ref{fig:mom}(a)) and 
pions (Fig.~\ref{fig:mom}(c)).
The dip in the $\cos \theta$ distribution of the muon (Fig.~\ref{fig:cos}(b))
is due to an efficiency drop at the overlapping region 
between the barrel and endcap KLM elements.

\begin{figure*}[t]
\centerline{\resizebox{5cm}{5cm}{\includegraphics{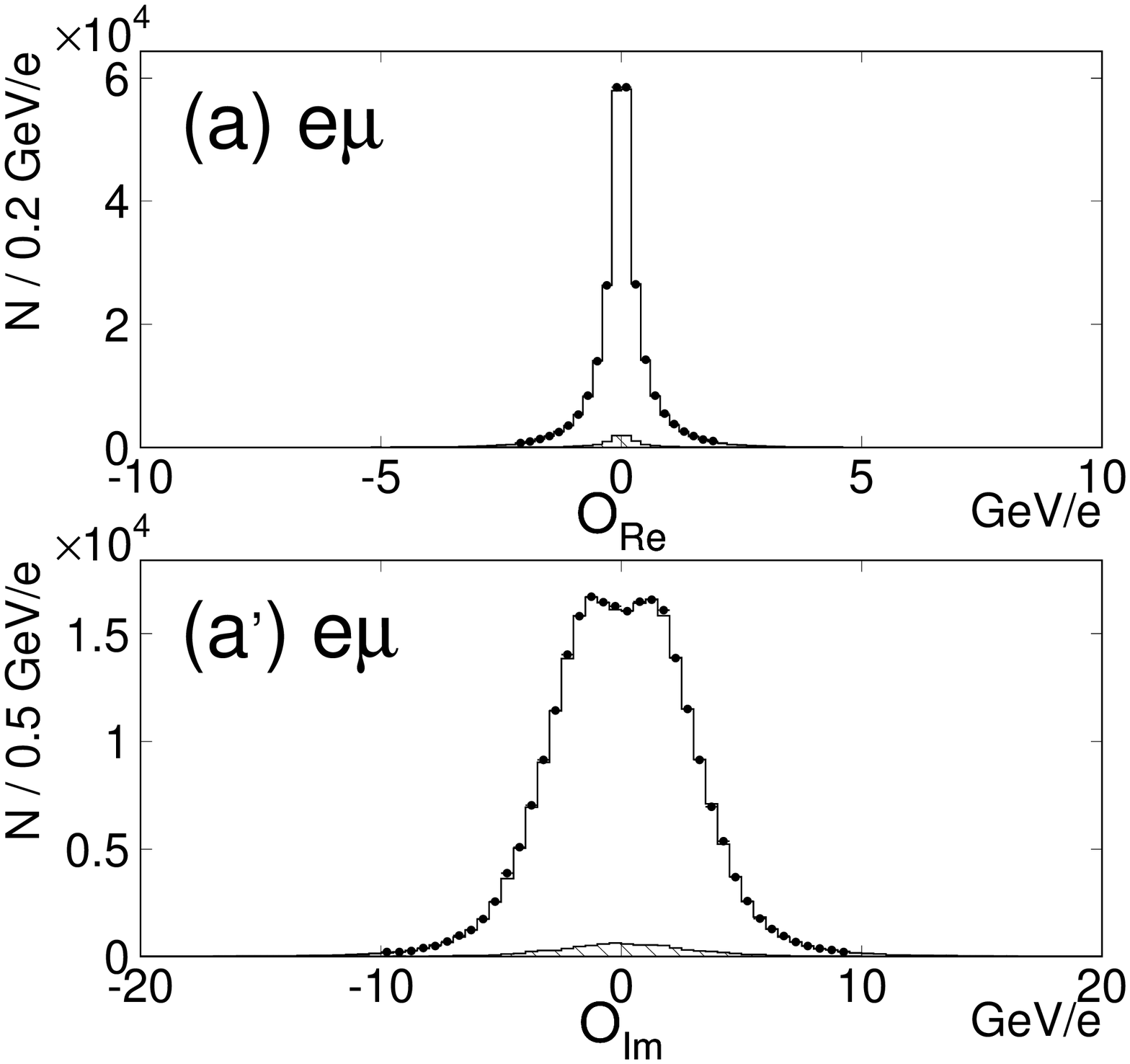}}
            \resizebox{5cm}{5cm}{\includegraphics{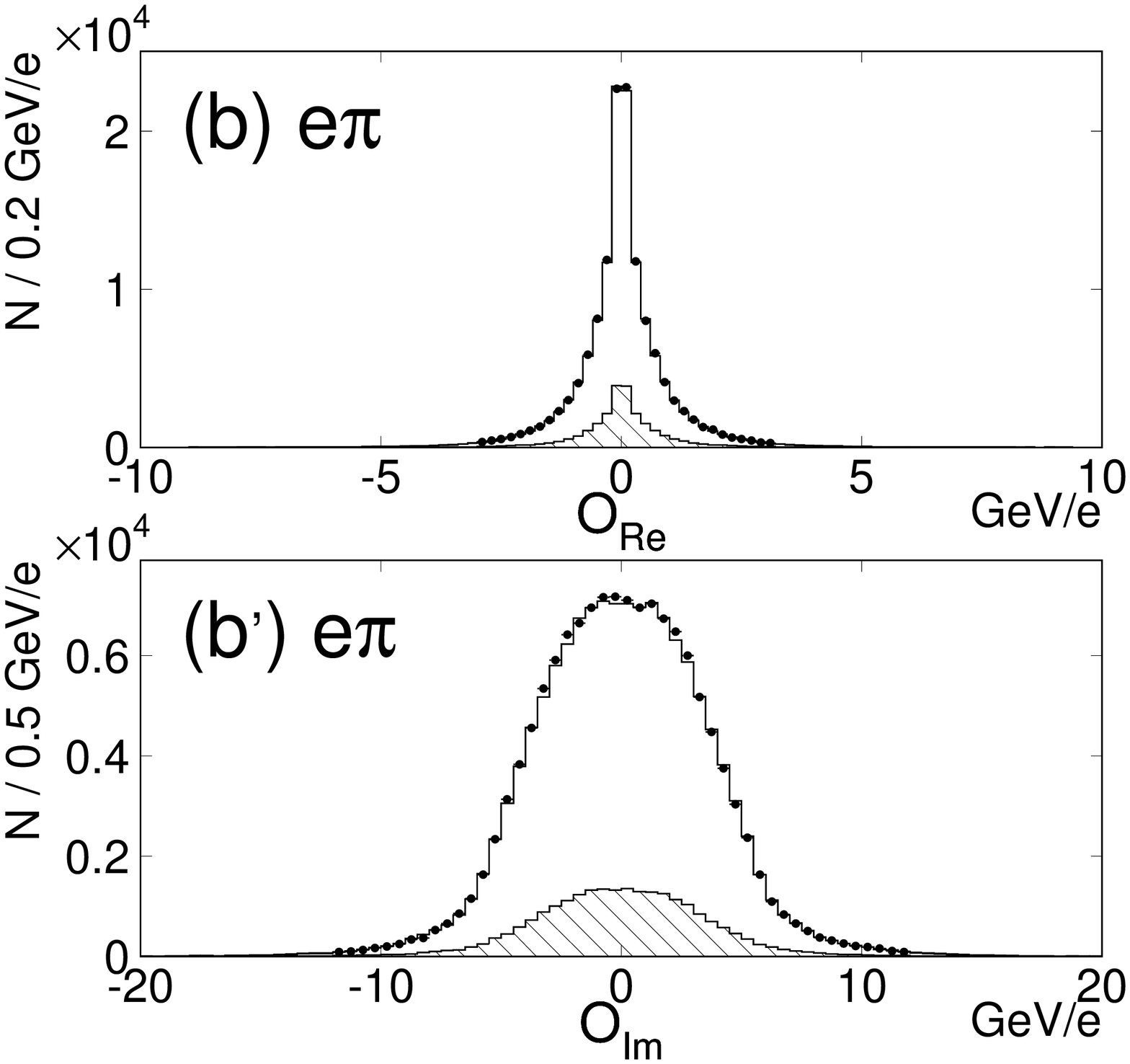}}
            \resizebox{5cm}{5cm}{\includegraphics{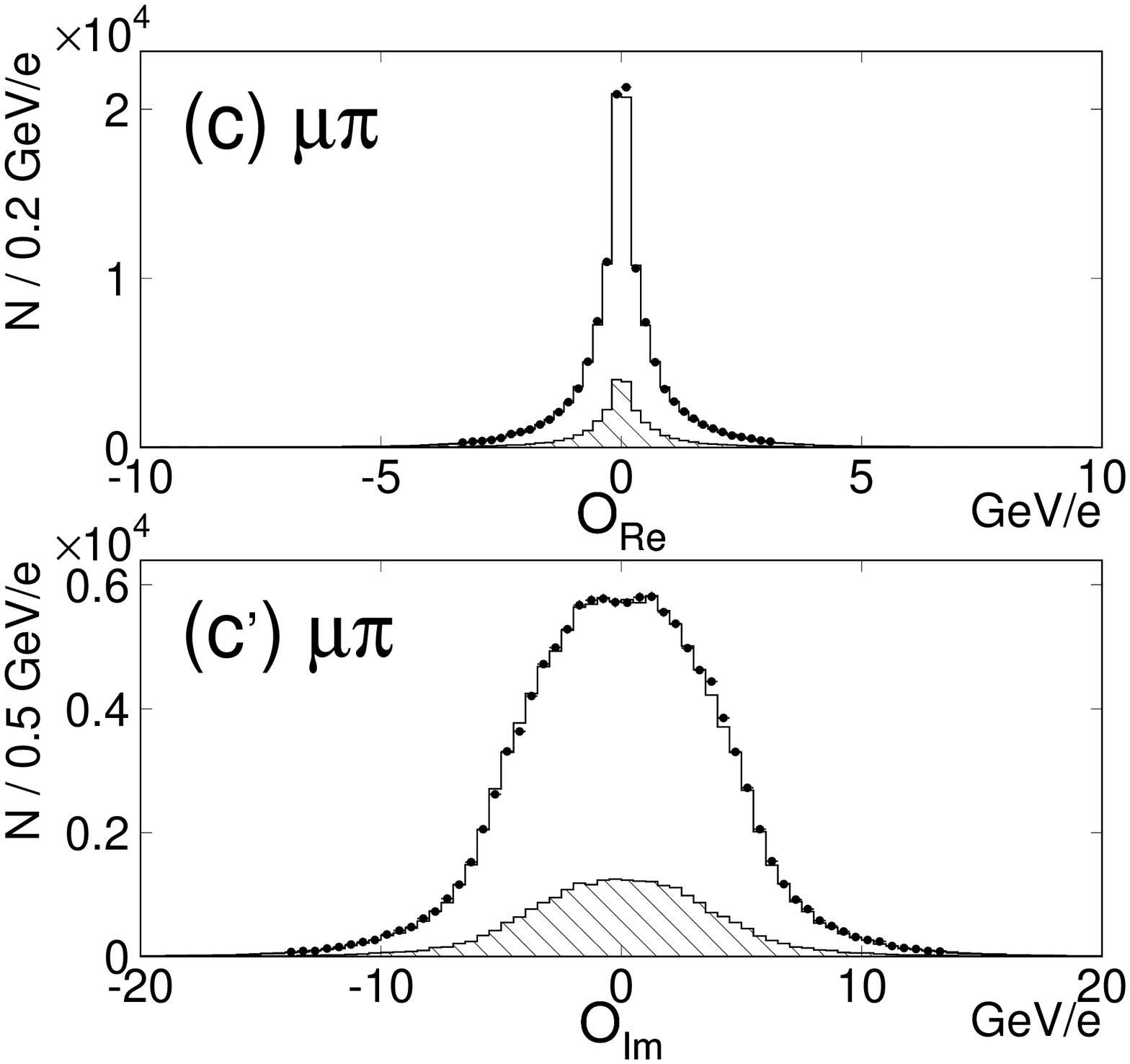}}}
\centerline{\resizebox{5cm}{5cm}{\includegraphics{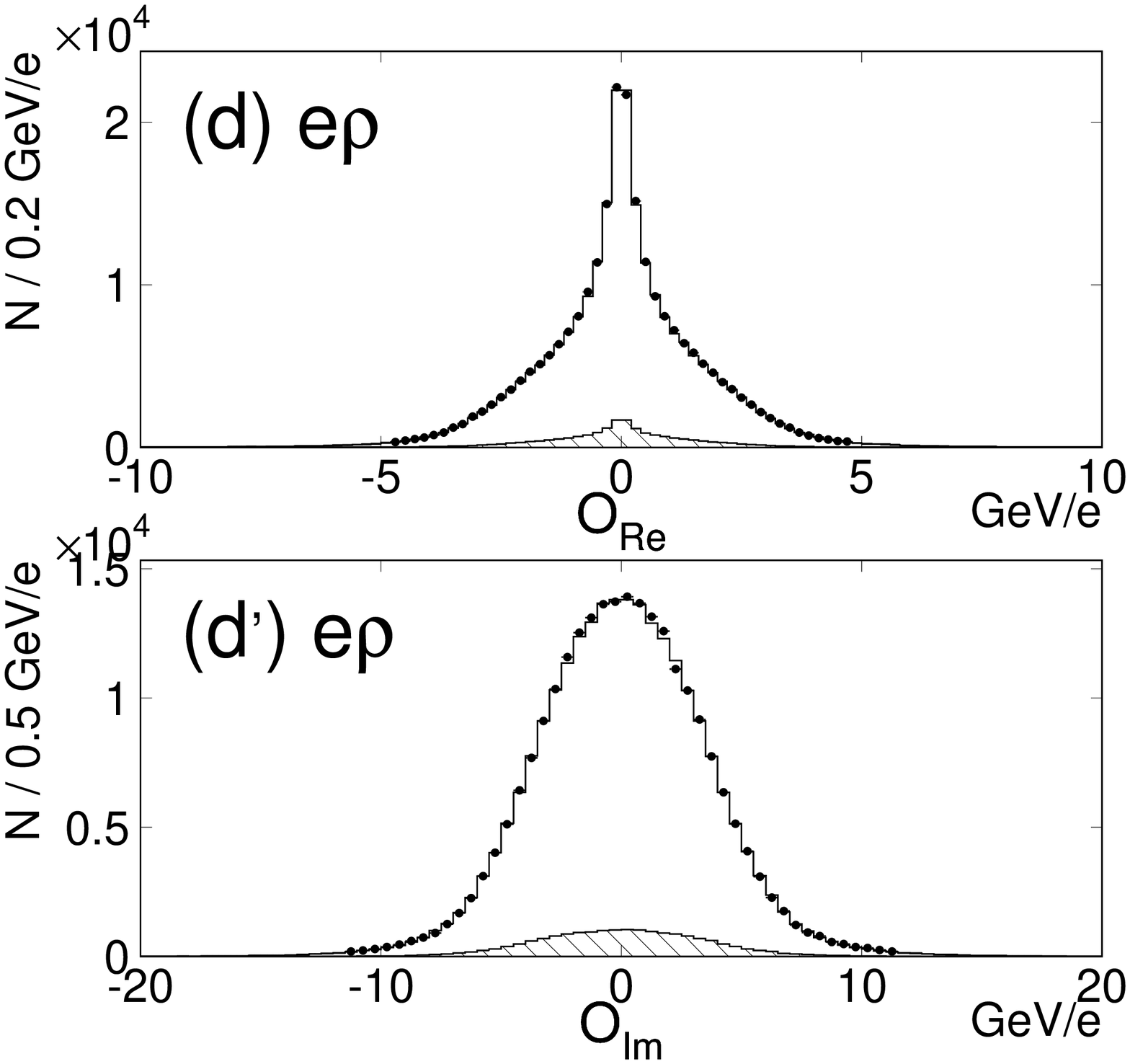}}
            \resizebox{5cm}{5cm}{\includegraphics{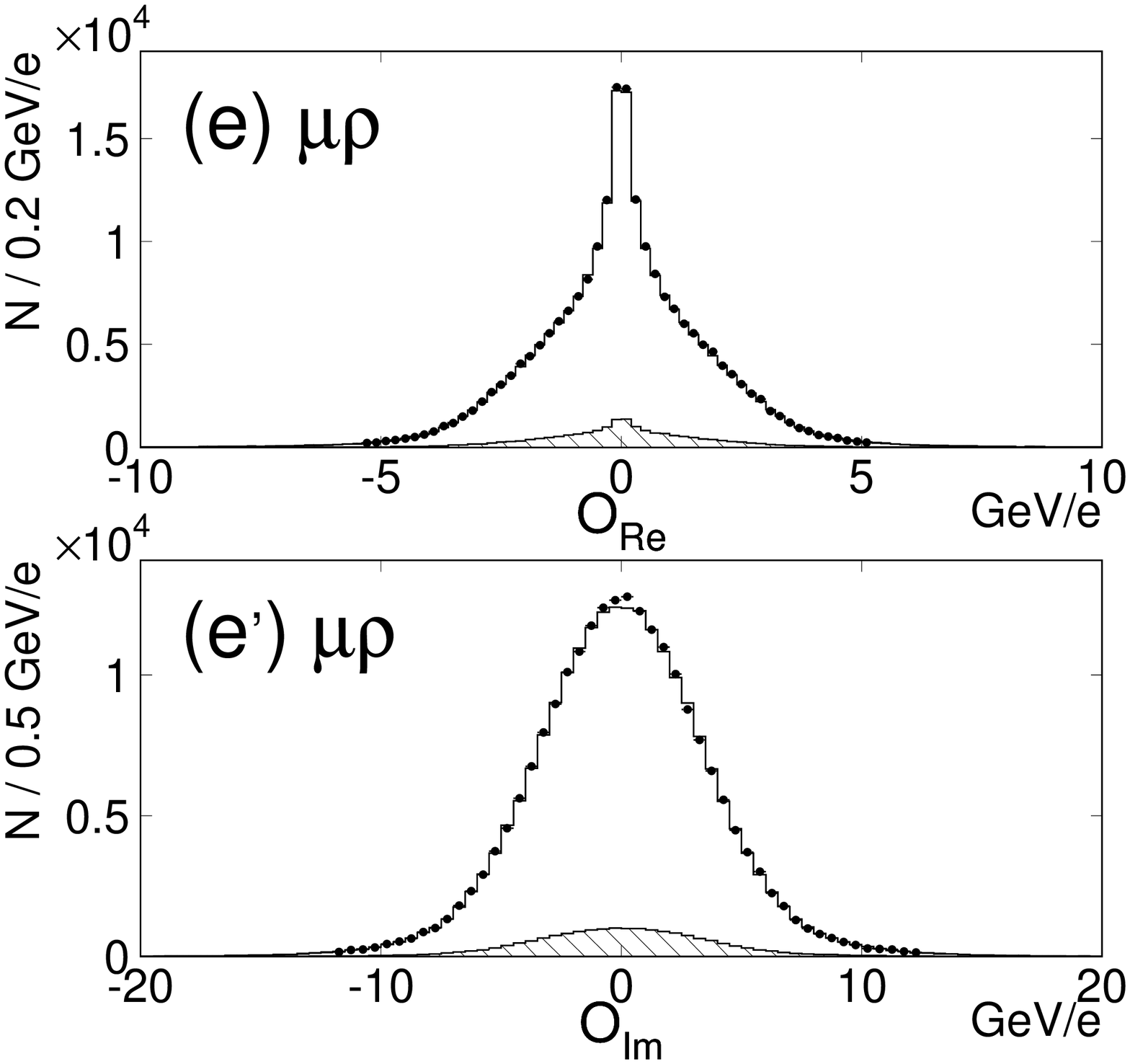}}
            \resizebox{5cm}{5cm}{\includegraphics{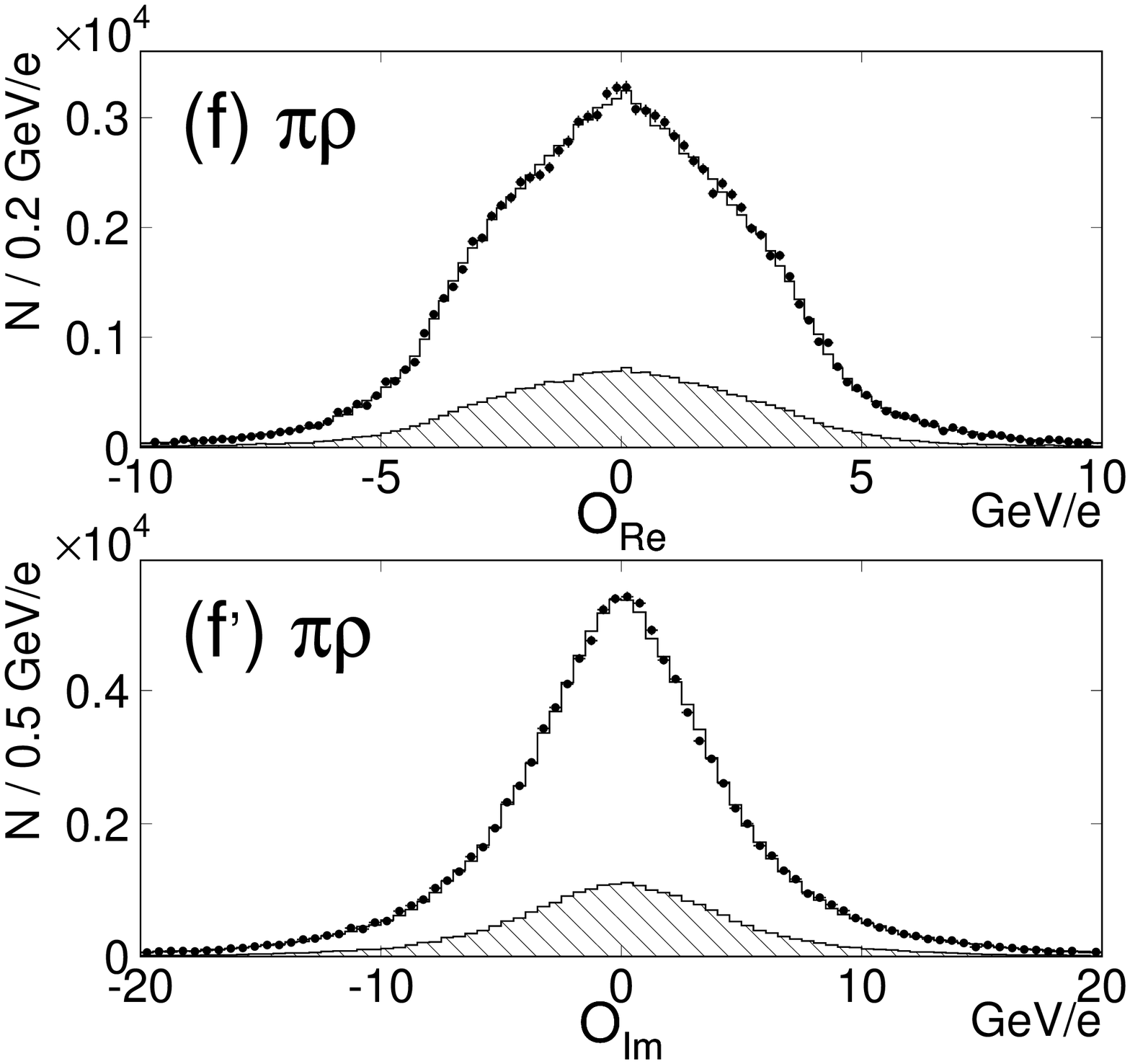}}}
\centerline{\resizebox{5cm}{5cm}{\includegraphics{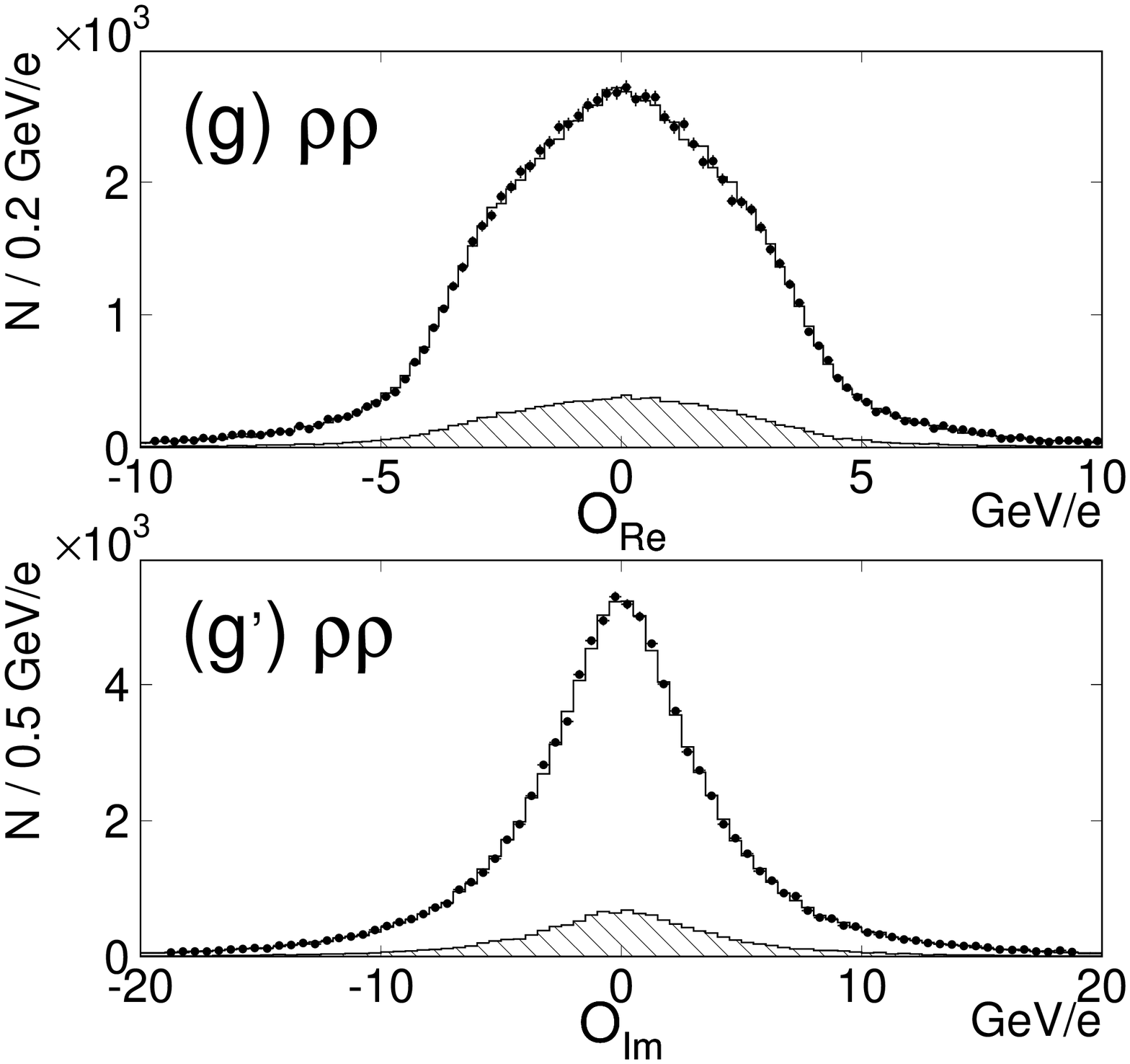}}
            \resizebox{5cm}{5cm}{\includegraphics{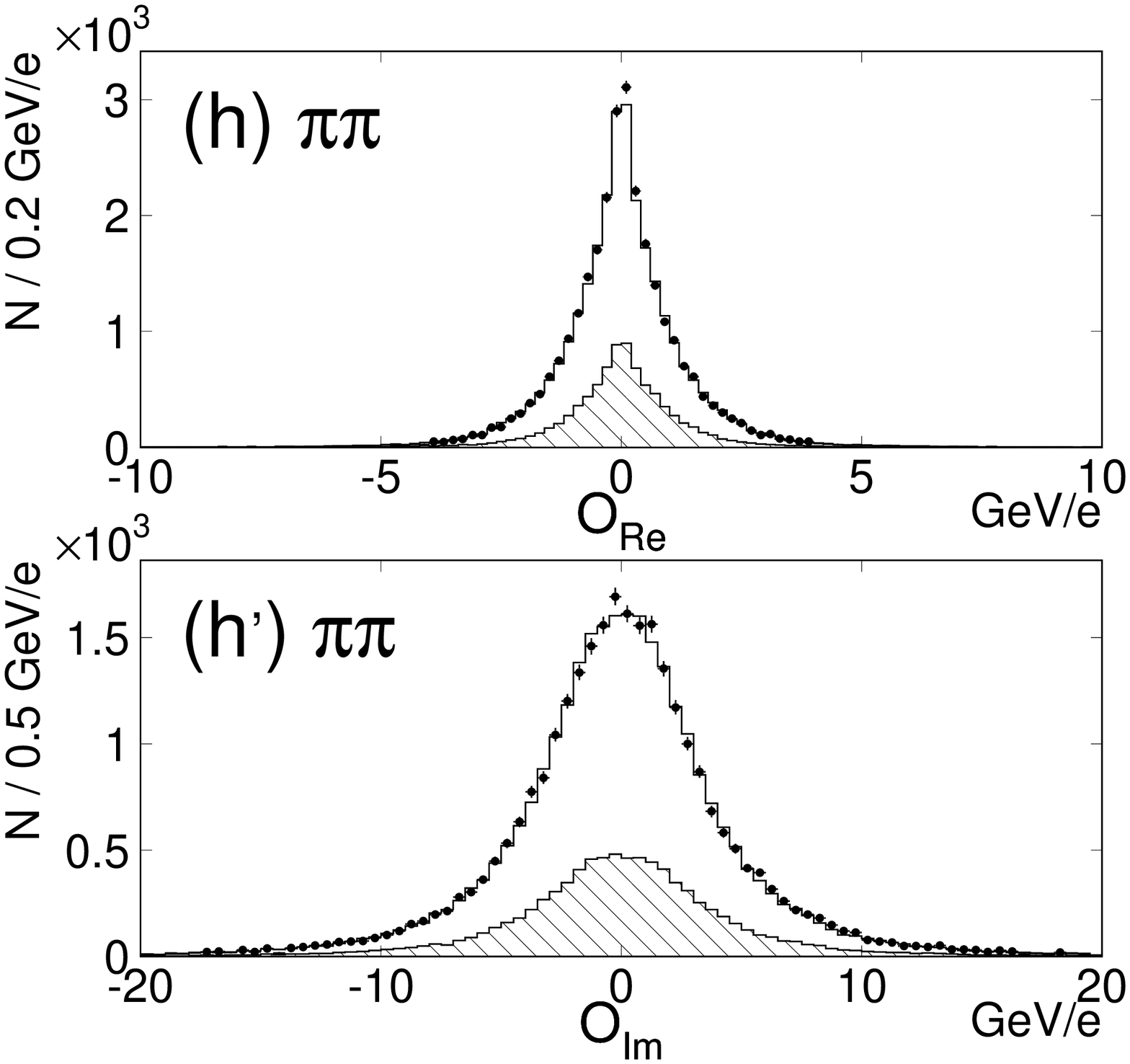}}
            \hspace*{5cm} }
\renewcommand {\baselinestretch}{0.75}
\caption{\small Distributions of the optimal observables 
${\cal O}_{Re}$ and ${\cal O}_{Im}$ for each mode. 
The upper figure of each mode is for ${\cal O}_{Re}$ 
and lower figure is for ${\cal O}_{Im}$.
The closed circles are the experimental data and the histogram is the MC 
expectation with $d_\tau=0$, normalized to the number of entries.
The hatched histogram is the background distribution evaluated by MC.}
\label{fig:Odistribution}
\end{figure*}

The resultant ${\cal O}_{Re}$ and ${\cal O}_{Im}$ distributions
are shown in Fig.~\ref{fig:Odistribution} along
with those obtained from MC simulation with $d_\tau = 0$.
Good agreement is found between the experimental data
and the MC samples.
The ratio of the data to MC shows flat distributions around 1.0,
although they are not shown here.

\section{Extraction of $d_\tau$} \label{sec:extraction}

In order to extract the $d_\tau$ value from the observable 
using Eq.~(\ref{eq:relation1}), we have to know the coefficient $a$ and 
the offset $b$.
In the ARGUS analysis~\cite{ref:ARGUS} which also used the optimal observable
method, the first term
of Eq.~(\ref{eq:obs}) was assumed to be negligible
because of the property of the CP violating term 
under the ideal detection hypothesis
$\int {\cal{M}}^2_{Re} d\phi = 0$.
The value of $d_{\tau}$ was thus obtained as 
the ratio of the observable's mean to the second term,
$Re(d_{\tau})=\langle{\cal O}_{Re}\rangle/\langle{\cal O}^2_{Re}\rangle$.  
Experimentally, the detector acceptance $\eta$ affects the
means of the observables as
\begin{equation}
\langle{\cal {O}}_{Re}\rangle \propto
\int \eta(\phi) {\cal O}_{Re} {\cal M}^2_{\rm prod} d\phi.
\end{equation}
Similar expressions are obtained for the imaginary part.
This means that the first term of Eq.~(\ref{eq:obs}) is not necessarily zero
and the coefficient may differ from $\langle{\cal O}^2_{Re}\rangle$
when the detector acceptance is taken into account.
In their study, the acceptance effect produced 
the largest systematic uncertainty, on the order of $10^{-16} e$cm.

\begin{figure}[tb]
\centerline{\resizebox{7cm}{3.5cm}{\includegraphics{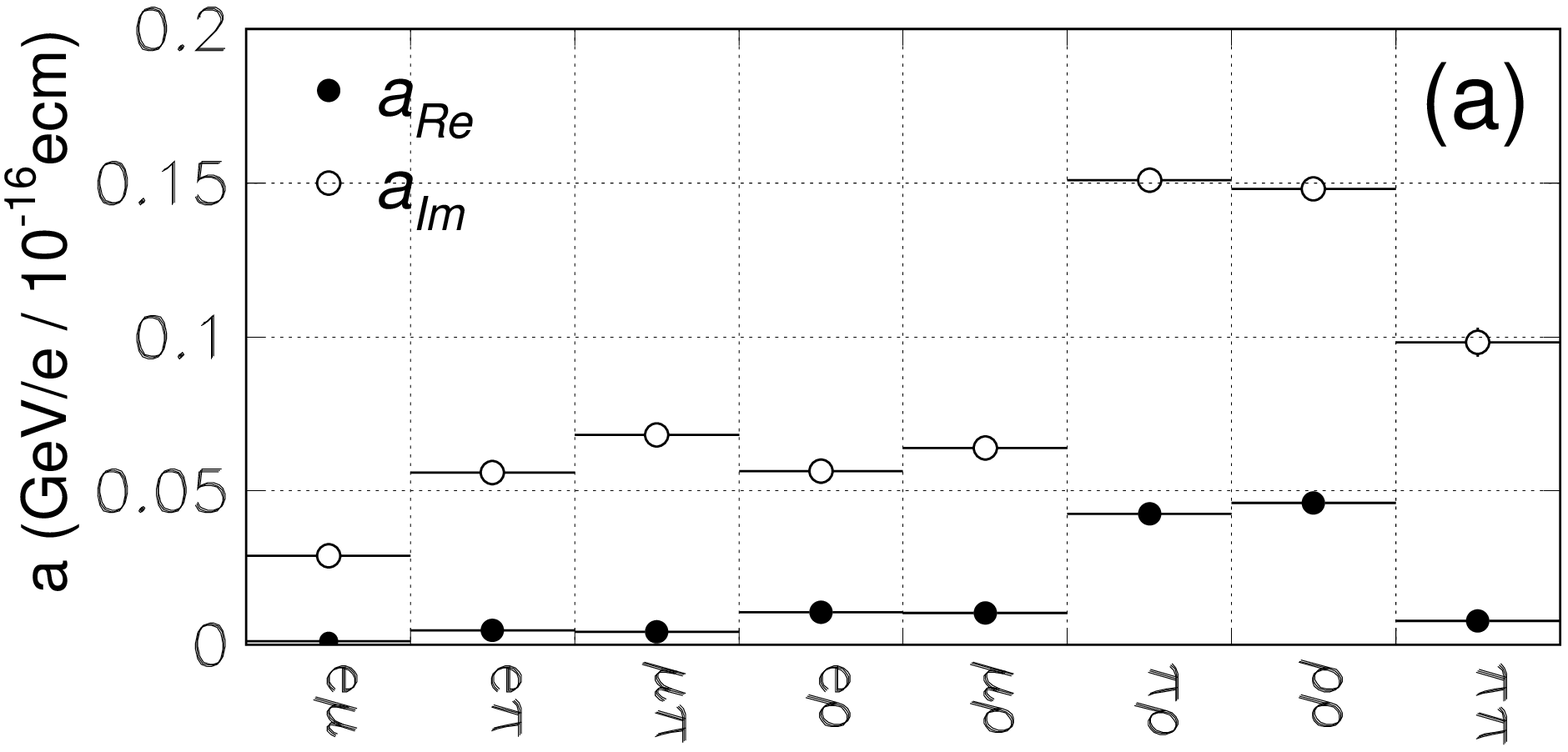}}}
\centerline{\resizebox{7cm}{3.5cm}{\includegraphics{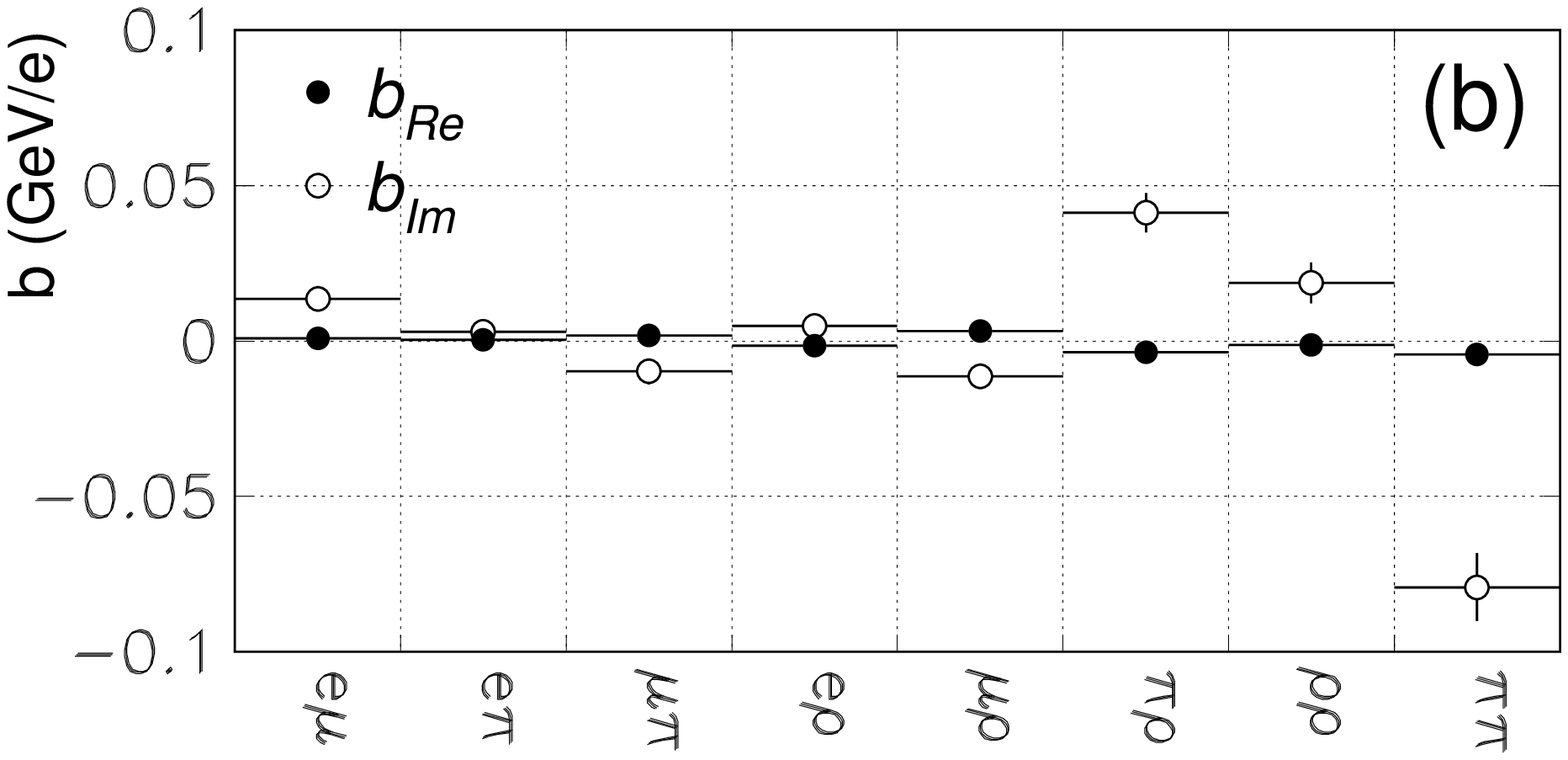}}}
\renewcommand {\baselinestretch}{0.75}
\caption{\small Sensitivity $a$ (a) and 
offset $b$ (b) for each mode from MC. 
The closed circles show the parameters for $Re(d_\tau)$
and the open circles show the parameters for $Im(d_\tau)$.
The errors are due to MC statistics.
}
\label{fig:sensitivity}
\end{figure}

In order to reduce this systematic effect,
we use both parameters $a$ and $b$ extracted from the correlation between
$\langle {\cal O}_{Re} \rangle$($\langle {\cal O}_{Im} \rangle$)
and $Re(d_\tau)$($Im(d_\tau)$)
obtained by a full MC including acceptance effects.
\begin{figure}[tb]
\centerline{\resizebox{7cm}{7cm}{\includegraphics{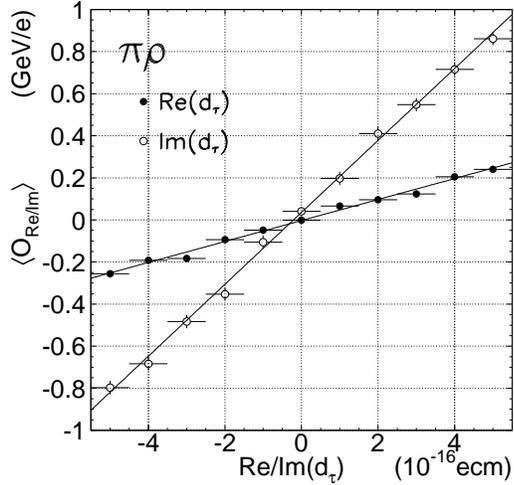}}}
\renewcommand {\baselinestretch}{0.75}
\caption{\small $Re(d_\tau)(Im(d_\tau))$ dependence of the mean of the 
observable $\langle{\cal O}_{Re}\rangle$($\langle{\cal O}_{Im}\rangle$)
for the $\pi\rho$ mode.
The closed circles show the dependence for $Re(d_\tau)$ and 
the open circles show the dependence for $Im(d_\tau)$.
The lines show the fitted linear functions.}
\label{fig:correlation}
\end{figure}
An example of the correlation between 
$\langle {\cal O}_{Re} \rangle$($\langle {\cal O}_{Im} \rangle$) 
and $Re(d_\tau)$($Im(d_\tau)$) 
is shown in Fig.~\ref{fig:correlation}. Each point
is obtained from MC with detector simulation and event selection. 
By fitting the correlation plot with Eq.~(\ref{eq:relation1}), the parameters 
$a$ and $b$ are obtained.
The misidentified background from $\tau$-pair events
shows some dependence on $d_\tau$, 
because the spin direction is correlated with the momenta of
the final state particles.
Therefore, both the coefficient $a$ and offset $b$
are corrected with the parameters obtained from the MC misidentified background.
The resultant coefficients and offsets are shown
in Fig.~\ref{fig:sensitivity}. 

As can be seen from the values of the coefficient $a$
showing the sensitivity to $d_\tau$,
the $\pi\rho$ and $\rho\rho$ modes have the highest sensitivities,  
while the $\pi\pi$ mode yields a somewhat lower value.
This is because both $\Vec{S}$ and $\Vec{k}$ in the $\pi\pi$ mode
are inferred only from the $\pi$ momentum vector, so that taking the
mean value of two possible solutions with equal weights significantly
reduces the achievable sensitivity. 
On the other hand, the momentum sum and angular distribution of 
$\rho \to \pi\pi^0$ decay in the $\pi\rho$ and $\rho\rho$ modes 
provide different weights 
on the possible solutions, and therefore there is less reduction 
in sensitivity.
Other modes, which include leptons, have low sensitivities
because of a lack of information about the $\tau^\pm$ directions and 
spin orientations due to the additional missing neutrinos. 

Non-zero offsets $b_{Im}$ are seen for the imaginary part,
due to the forward/backward asymmetry in the acceptance of the detector.

\section{Systematic uncertainties}

\begin{table*}[htb]
\renewcommand {\baselinestretch}{0.75}
 \caption{Systematic errors for $Re(d_\tau)$ and $Im(d_\tau)$
in units of $10^{-16}e{\rm cm}$.}
 \label{table:Syserror}
 \begin{tabular}{lcccccccc}
  \hline
$Re(d_\tau)$ &$e\mu$&$e\pi$&$\mu\pi$&$e\rho$&$\mu\rho$&$\pi\rho$&$\rho\rho$&$\pi\pi$\\
  \hline
Mismatch of distribution & 0.80 & 0.58 & 0.70 & 0.11 & 0.15 & 0.21 & 0.16 & 0.06 \\
Charge asymmetry         & 0.00 & 0.01 & 0.01 & 0.01 & 0.01 & 0.01 & -    & -    \\
Background variation     & 0.43 & 0.12 & 0.07 & 0.07 & 0.08 & 0.03 & 0.04 & 0.05 \\
Momentum reconstruction  & 0.16 & 0.09 & 0.24 & 0.04 & 0.06 & 0.06 & 0.04 & 0.45 \\
Detector alignment       & 0.02 & 0.02 & 0.01 & 0.00 & 0.01 & 0.01 & 0.02 & 0.03 \\
  \hline
Total                    & 0.92 & 0.60 & 0.74 & 0.14 & 0.18 & 0.22 & 0.17 & 0.46 \\
  \hline \hline
$Im(d_\tau)$ &$e\mu$&$e\pi$&$\mu\pi$&$e\rho$&$\mu\rho$&$\pi\rho$&$\rho\rho$&$\pi\pi$\\
  \hline
Mismatch of distribution & 0.43 & 0.02 & 0.05 & 0.12 & 0.01 & 0.05 & 0.10 & 0.41 \\
Charge asymmetry         & 0.13 & 0.44 & 0.43 & 0.02 & 0.09 & 0.15 & -    & -    \\
Background variation     & 0.08 & 0.07 & 0.02 & 0.01 & 0.03 & 0.02 & 0.03 & 0.06 \\
Momentum reconstruction  & 0.03 & 0.03 & 0.06 & 0.00 & 0.02 & 0.02 & 0.04 & 0.04 \\
Detector alignment       & 0.01 & 0.03 & 0.02 & 0.01 & 0.01 & 0.02 & 0.01 & 0.05 \\
  \hline
Total                    & 0.46 & 0.45 & 0.44 & 0.13 & 0.10 & 0.16 & 0.11 & 0.42 \\
  \hline
 \end{tabular}
\end{table*}

Although in general the MC simulation reproduces the observed 
kinematic distributions well, the disagreement
seen in Figs.~\ref{fig:mom} and \ref{fig:cos} causes an error
dominating the systematic uncertainty.
Its uncertainty with respect to $d_\tau$ was studied by reweighting 
the MC distributions by the ratio of data to MC.
The second significant uncertainty originates from possible charge asymmetry
in the detection efficiency. 
The ratio of yields, $N(\alpha^+\beta^-)/N(\alpha^-\beta^+)$,
for data and MC agrees within 1\% accuracy, 
where $\alpha$ and $\beta$ are the relevant charged particles
from the $\tau$ decays. 
This systematic uncertainty was evaluated by varying the detection 
efficiency by $\pm1$\%. 
The effect is of the same size as the statistical error for $Im(d_{\tau})$,
while it is negligible for $Re(d_{\tau})$. 
The backgrounds provide additional systematic errors for $d_\tau$ 
because the parameters $a$ and $b$ are corrected for
the background contributions,
which are assessed by varying the background rate.
The influence of the momentum reconstruction of charged particles and
photons was also checked by applying scaling factors 
corresponding to the difference between the data and MC.
In order to examine a possible asymmetry arising from 
the alignment of the tracking devices,
we measured the differences of the polar angles, 
${\it \Delta} \theta = \theta^+_{\rm CM} - \theta^-_{\rm CM}$,
and the azimuthal angles, 
${\it \Delta} \phi = \phi^+_{\rm CM} - \phi^-_{\rm CM}$,
between two tracks in $e^+e^- \to \mu^+\mu^-$ events,
and found a small deviation from a back-to-back topological alignment
in each direction as 
${\it \Delta} \theta = 1.48$ mrad and ${\it \Delta} \phi = 0.36$ mrad.
By applying an artificial angular deviation of this magnitude 
to one of the charged tracks, 
the residual value of the observables is calculated to be 
negligible compared with the other errors.
The systematic errors are listed in Table~\ref{table:Syserror}.

\section{Result}

The values of $d_\tau$ extracted using Eq.~(\ref{eq:relation1}) are listed in 
Table~\ref{table:EDMresult} along with the statistical and systematic errors.
All results are consistent with zero within the errors.

\begin{table}[tb]
\renewcommand {\baselinestretch}{0.75}
 \caption{Results for the electric dipole form factor. 
The first error is statistical and the second is systematic.}
 \label{table:EDMresult}
 \begin{tabular}{ccc}
  \hline
             & $Re(d_\tau)~(10^{-16}e{\rm cm})$&$Im(d_\tau)~(10^{-16}e{\rm cm})$\\
  \hline
      $e\mu$ & $~~2.25 \pm 1.26 \pm 0.92$ & $ -0.41 \pm 0.22 \pm 0.46$ \\
      $e\pi$ & $~~0.43 \pm 0.64 \pm 0.60$ & $ -0.22 \pm 0.19 \pm 0.45$ \\
    $\mu\pi$ & $ -0.41 \pm 0.87 \pm 0.74$ & $~~0.15 \pm 0.19 \pm 0.44$ \\
     $e\rho$ & $~~0.00 \pm 0.36 \pm 0.14$ & $ -0.01 \pm 0.14 \pm 0.13$ \\
   $\mu\rho$ & $~~0.04 \pm 0.42 \pm 0.18$ & $ -0.02 \pm 0.14 \pm 0.10$ \\
   $\pi\rho$ & $~~0.34 \pm 0.25 \pm 0.22$ & $ -0.22 \pm 0.13 \pm 0.16$ \\
  $\rho\rho$ & $ -0.08 \pm 0.25 \pm 0.17$ & $ -0.12 \pm 0.14 \pm 0.11$ \\
    $\pi\pi$ & $~~0.42 \pm 1.17 \pm 0.46$ & $~~0.24 \pm 0.34 \pm 0.42$ \\
  \hline
  All & $0.115 \pm 0.170$          & $-0.083 \pm 0.086$ \\
  \hline
 \end{tabular}
\end{table}

Finally, we obtain mean values for $Re(d_\tau)$ and $Im(d_\tau)$
over the eight different $\tau^+\tau^-$ modes weighted by quadratically
summed statistical and systematic errors.
The resultant preliminary result for the electric dipole form factors are 
\begin{eqnarray}
 Re(d_\tau) &=& ( 1.15 \pm 1.70 ) \times 10^{-17} e{\rm cm}, \\
 Im(d_\tau) &=& ( -0.83 \pm 0.86 ) \times 10^{-17} e{\rm cm},
\end{eqnarray}
with corresponding 95\% confidence limits
\begin{eqnarray}
-2.2 < Re(d_\tau) < 4.5 ~~~(10^{-17} e{\rm cm}), \\
-2.5 < Im(d_\tau) < 0.8 ~~~(10^{-17} e{\rm cm}).
\end{eqnarray}

This investigation has improved the sensitivity to the $\tau$ lepton's
electric dipole form factor by a factor of more than 10 compared to 
previous measurements. 

\smallskip
\bigskip
\noindent
{\bf Acknowledgements}
\smallskip

We would like to thank Professors K.~Hagiwara, O.~Nachtmann, and Z.~W\c{a}s 
for their constructive advices and many helpful discussions.
We also wish to thank the KEKB accelerator group for the excellent
operation of the KEKB accelerator.
We gratefully acknowledge the support from the Ministry of Education,
Culture, Sports, Science, and Technology of Japan,
Grant-in-Aid for JSPS Fellows 01655 2001,
and the Japan Society for the Promotion of Science;
the Australian Research Council
and the Australian Department of Industry, Science and Resources;
the National Science Foundation of China under contract No.~10175071;
the Department of Science and Technology of India;
the BK21 program of the Ministry of Education of Korea
and the CHEP SRC program of the Korea Science and Engineering Foundation;
the Polish State Committee for Scientific Research
under contract No.~2P03B 17017;
the Ministry of Science and Technology of the Russian Federation;
the Ministry of Education, Science and Sport of the Republic of Slovenia;
the National Science Council and the Ministry of Education of Taiwan;
and the U.S.\ Department of Energy.


\begin{thebibliography}{99}
\bibitem{ref:BCPV1} K.~Abe et al. (Belle Collaboration),
Phys. Rev. Lett. {\bf 87} (2001) 091802.
\bibitem{ref:BCPV2} B.~Aubert et al. (BaBar Collaboration),
Phys. Rev. Lett. {\bf 87} (2001) 091801.
\bibitem{ref:th1} T.~Huang, W.~Lu, and Z.~Tao, 
Phys. Rev. {\bf D55} (1997) 1643.
\bibitem{ref:th2} W.~Bernreuther, A.~Brandenburg, and P.~Overmann,
Phys. Lett. {\bf B391} (1997) 413; ibid., {\bf B412} (1997) 425.
\bibitem{ref:L3} M.~Acciarri et al. (L3 Collaboration),
Phys. Lett. {\bf B434} (1998) 169.
\bibitem{ref:OPAL} K.~Ackerstaff et al. (OPAL Collaboration),
Phys. Lett. {\bf B431} (1998) 188.
\bibitem{ref:ARGUS} H.~Albrecht et al. (ARGUS Collaboration),
Phys. Lett. {\bf B485} (2000) 37.
\bibitem{ref:EDM} W.~Bernreuther, O.~Nachtmann, and P.~Overmann, 
Phys. Rev. {\bf D48} (1993) 78.
\bibitem{ref:Optimal} D.~Atwood and A.~Soni, Phys. Rev. {\bf D45} (1992) 2405.
\bibitem{ref:Belle} A.~Abashian et al. (Belle Collaboration),
Nucl. Instr. and Meth. A {\bf 479} (2002) 117.
\bibitem{ref:KEKB} E.~Kikutani ed., KEK Preprint 2001-157 (2001), 
 to appear in Nucl. Instr. and Meth. A.
\bibitem{ref:KORALB} KORALB(v.2.4)/TAUOLA(v.2.6): S.~Jadach and Z.~W\c{a}s, 
Comp. Phys. Commun. {\bf 85} (1995) 453; ibid., {\bf 64} (1991) 267;
ibid., {\bf 36} (1985) 191; S.~Jadach, Z.~W\c{a}s, R.~Decker, and J.H.~K\"uhn,
Comp. Phys. Commun. {\bf 64} (1991) 275; ibid., {\bf 70} (1992) 69;
ibid., {\bf 76} (1993) 361.
\bibitem{ref:eID} K.~Hanagaki et al.,
Nucl. Instr. and Meth. A {\bf 485} (2002) 490.
\end{thebibliography}
\end{document}